\documentclass[aps,prl,reprint,twocolumn,showpacs,superscriptaddress,floatfix]{revtex4}
\usepackage{amsmath}
\usepackage[ansinew]{inputenc}
\usepackage[english]{babel}
\usepackage{graphicx}
\usepackage{mathrsfs}

\newcommand{\Li}{\ensuremath{^7\text{Li}^+} }

\newcommand{\DSE}{\ensuremath{^3\text{S}_1} }
\newcommand{\DPZ}{\ensuremath{^3\text{P}_2} }

\begin{document}

\title{Test of Time Dilation Using Stored Li$^+$ Ions as Clocks at Relativistic Speed}


\author{Benjamin Botermann}
\affiliation{Johannes Gutenberg-Universit\"at Mainz, Institut f\"ur
Kernchemie, 55128 Mainz, Germany}
\author{Dennis Bing}
\affiliation{Max-Planck-Institut f\"ur Kernphysik, 69117 Heidelberg,
Germany}
\author{Christopher Geppert}
\affiliation{Johannes Gutenberg-Universit\"at Mainz, Institut f\"ur
Kernchemie, 55128 Mainz, Germany}
\affiliation{GSI Helmholtzzentrum
f\"ur Schwerionenforschung, 64291 Darmstadt, Germany}
\affiliation{Helmholtzinstitut Mainz, 55128 Mainz, Germany}
\author{Gerald Gwinner}
\affiliation{Department of Physics and Astronomy, University of Manitoba, Winnipeg, Manitoba R3T 2N2, Canada}
\author{Theodor W. H\"ansch}
\affiliation{Max-Planck-Institut f\"ur Quantenoptik, 85748 Garching,
Germany}
\author{Gerhard Huber}
\affiliation{Johannes Gutenberg-Universit\"at Mainz, Institut f\"ur
Physik, 55128 Mainz, Germany}
\author{Sergei Karpuk}
\affiliation{Johannes Gutenberg-Universit\"at Mainz, Institut f\"ur
Physik, 55128 Mainz, Germany}
\author{Andreas Krieger}
\affiliation{Johannes Gutenberg-Universit\"at Mainz, Institut f\"ur
Kernchemie, 55128 Mainz, Germany}
\author{Thomas K\"uhl}
\affiliation{GSI Helmholtzzentrum f\"ur Schwerionenforschung, 64291 Darmstadt, Germany}
\author{Wilfried N\"ortersh\"auser}
\affiliation{Johannes Gutenberg-Universit\"at Mainz, Institut f\"ur
Kernchemie, 55128 Mainz, Germany} \affiliation{GSI Helmholtzzentrum
f\"ur Schwerionenforschung, 64291 Darmstadt, Germany} \affiliation{TU
Darmstadt, Institut f\"ur Kernphysik, 64289 Darmstadt, Germany}
\author{Christian Novotny}
\affiliation{Helmholtzinstitut Mainz, 55128 Mainz, Germany}
\author{Sascha Reinhardt}
\affiliation{Max-Planck-Institut f\"ur Quantenoptik, 85748 Garching,
Germany}
\author{Rodolfo S\'anchez}
\affiliation{Helmholtzinstitut Mainz, 55128 Mainz, Germany}
\author{Dirk Schwalm}
\affiliation{Max-Planck-Institut f\"ur Kernphysik, 69117 Heidelberg,
Germany}
\author{Thomas St\"ohlker}
\affiliation{GSI Helmholtzzentrum f\"ur Schwerionenforschung, 64291 Darmstadt, Germany}
\author{Andreas Wolf}
\affiliation{Max-Planck-Institut f\"ur Kernphysik, 69117 Heidelberg,
Germany}
\author{Guido Saathoff}
\affiliation{Max-Planck-Institut f\"ur Quantenoptik, 85748 Garching,
Germany}

\date{\today}

\begin{abstract}

We present  the {concluding} result from an Ives-Stilwell-type time
dilation experiment using \Li ions confined at a velocity of
$\beta=v/c = 0.338$ in the storage ring ESR at Darmstadt. A
$\Lambda$-type three-level system within the hyperfine structure of
the \Li \DSE $\! \rightarrow\!$ \DPZ line is driven by two laser
beams aligned parallel and antiparallel relative to the ion beam.
The lasers' Doppler shifted frequencies required for resonance are
measured with an accuracy of $ < 4 \times 10^{-9}$ using
optical-optical double resonance spectroscopy. This allows us to verify
the Special Relativity relation between the time dilation factor
$\gamma$ and the velocity $\beta$, $\gamma \sqrt{1-\beta^2} = 1$ to
within ${\pm}2.3 \times 10^{-9}$ {at this velocity}. The result,
which is singled out by a high boost velocity $\beta$, is also
interpreted within Lorentz Invariance violating test theories.

\end{abstract}

\maketitle


With Special Relativity (SR), Local Lorentz Invariance (LI)
has been established as one of the corner stones of all currently
accepted theories describing nature on a fundamental level. Although
empirically well established, the fundamental role of this space
time symmetry in physics has accounted for incessant experimental
tests with ever increasing scrutiny~\cite{Mat05}. Interest in LI
tests have been further boosted by the search for a theory
reconciling quantum theory with general relativity, as many attempts
for such a quantum gravity explicitly allow Lorentz
violation~\cite{Kos89,Ame13,Rov08}, making it a potential
discriminatory experimental signature for the underlying theory.

Within the wealth of Lorentz invariance tests, Ives-Stilwell
(IS) experiments~\cite{Ive38}
stand out by using a large
experimentally prepared Lorentz boost, which neither depends on
sidereal variations nor on the assumption of any particular ({\it{ad hoc}}
chosen) reference frame~\cite{Gwi05}. It
directly verifies the time dilation factor $\gamma$, a salient consequence of
Special Relativity (SR) of epistemological and technological relevance, via
the
relativistic Doppler formula:
In SR the transition frequency $\nu_i$ of an
atom at rest in an inertial system $\tilde{S}$, which is moving with
a constant velocity $\beta=v/c$ in the laboratory system $S$, is
related to the frequency $\nu$ measured by an observer at rest in
$S$ by
\begin{equation} \label{eq:OLT}
\nu_i = \nu \gamma (1-\beta \cos \vartheta),
\end{equation}
where $\vartheta$ is the observation angle with respect to the atom
velocity $\beta$ and $\gamma=1/\sqrt{1-\beta^2}$. Ives and Stilwell
were the first who actually showed the square root dependence
of $\gamma$ on $(1-\beta^2)$ by measuring the wavelength of the
H$_\beta$ line emitted parallel and antiparallel to hydrogen canal
rays~\cite{Ive38}. We have implemented a modern version of this
principle by driving two transitions $\nu_1$ and $\nu_2$ in a
$^7$Li$^+$ ion moving at velocity $\beta$ with a copropagating
(parallel) ($p$) and a counterpropagating (antiparallel) ($a$) laser
beam. For resonance, the frequencies $\nu_p$ and $\nu_a$ of the two
laser beams need to obey Eq.~(\ref{eq:OLT}), which results in
\begin{equation}\label{eq:Doppler1}
 \frac{\nu_a \nu_p}{\nu_1 \nu_2} = \frac{1}{\gamma^2 (1-\beta^2)}  = 1,
\end{equation}
when compared to the corresponding rest-frame frequencies $\nu_1$
and $\nu_2$ measured in the laboratory system $S$. Note that the
$\beta$ independence of this relation and its contraction to one is
not only a consequence of the special velocity dependence of the
(kinematic) Lorentz factor $\gamma$ governing time dilation but
also relies on the validity of the relativity principle, namely that the atomic
transition frequencies and the phase velocity of electromagnetic
waves in vacuum are invariant under active (particle) Lorentz
transformation.

Eq.~(\ref{eq:Doppler1}) can be tested to high accuracy as it only
depends on how precisely the four laboratory frequencies can be
determined. While the boost velocity $\beta$ cancels out and thus
does not enter the error budget, the time dilation factor
$\gamma$, however, increases to lowest order quadratically with
$\beta$. Instead of quoting the IS observable
\begin{equation}\label{eq:Doppler2}
\epsilon(\beta) = \sqrt{\frac{\nu_a \nu_p}{\nu_1 \nu_2}} - 1
\end{equation}
it is therefore customary to use the reduced quantity $\alpha =
\epsilon(\beta)/\beta^2$ when comparing different experiments aiming
at a validation of Eq.~(\ref{eq:Doppler1}).
Since the first experiment by Ives and Stilwell~\cite{Ive38}, which
was performed at $\beta=0.005$ and resulted in $|\alpha| \leq
10^{-2}$, several IS experiments have been performed with ever
increasing sensitivity~\cite{MacArthur,McGowan,Gri94,Saa03,Rei07}.
So far, the most accurate
measurement
employed saturation spectroscopy on $^7$Li$^+$ ions stored at
$\beta=0.03$ and $\beta=0.064$ in the Heidelberg heavy ion storage
ring (TSR) and resulted in $|\alpha| \leq 8.4 \times
10^{-8}$~\cite{Rei07}.

The present Letter describes a measurement of $\epsilon$ employing
optical-optical double resonance  (OODR) spectroscopy on metastable
$^7$Li$^+$ ions stored at relativistic velocities of 33.8\%c in the
experimental storage ring ESR of the GSI Helmholtzzentrum at
Darmstadt. In a first version of this experiment~\cite{Nov09}, we
showed the feasibility of OODR spectroscopy for an IS test under the
harsh conditions of an accelerator laboratory. Here, we report on the
analysis of an improved set of OODR measurements taking full account
of their systematic uncertainties.  As explained below, OODR on a
$\Lambda$-type three-level transition yields higher signal-to-noise
which is decisive for the present experiment as well as for
sub-Doppler spectroscopy on thin ion beams in general. Also, OODR on
an appropriate $\Lambda$ transition with one transition frequency in
the infrared might allow us to mitigate the strong Doppler blueshift
keeping the necessary frequency for parallel excitation within the
reach of narrowband lasers. This way, sub-Doppler spectroscopy might
become feasible on ultrarelativistic beams of the upcoming FAIR
facility.

The final result derived for $\epsilon$, which constitutes the
hitherto most sensitive model-independent validation of
Eq.~\ref{eq:Doppler1}, will also be confronted with
Lorentz-violating test theories. Because of its large Lorentz boost, our
experiment might be especially sensitive to parameters of higher
mass dimension in the Standard Model Extension test
theory~\cite{Kos13}.


The \DSE $\! \rightarrow\!$ \DPZ transition of $^7$Li$^+$ is an
appropriate clock for Doppler shift experiments. The metastable \DSE
state lifetime of 50~s~\cite{Dra71,Kni80} in vacuum, although
reduced to $15 - 30$~s by collisions with residual gas atoms in the ESR,
is still sufficiently long for spectroscopy. The transition
wavelength of 548.5~nm is well within the optical region so that
both parallel and antiparallel probe light can be generated by
continuous wave lasers. The natural linewidth of 3.7~MHz is narrow
enough to reach sub-MHz accuracy, the domain where systematic
effects become dominant.
And finally, Li ions
can be accelerated and stored at high velocities and with excellent
beam qualities.

The $^7$Li$^+$ ions are generated in a Penning ion gauge 
(PIG) source and accelerated by the GSI accelerator facility to a final
energy of 58.6~MeV/u, which corresponds to a velocity of
$\beta=0.338$. The \Li ions are then transferred to the ESR where
electron cooling~\cite{Ste04} is employed to reduce the ion beam's
longitudinal
velocity spread to~$\delta v/v \approx 9 \times 10^{-6}$ (FWHM)
and its transverse divergence in the experimental section
to~$<100~\mu$rad. The metastable \DSE triplet ground state is weakly
populated  in the PIG source. After about 2~s of acceleration and
injection and $7 - 20\,$s of electron cooling, $10^7$ to $10^8$ ions are
stored in the ESR with an estimated fraction of less than 1~\% in
the metastable \DSE state.

The \Li transitions are excited by
lasers overlapping the ion beam parallel and antiparallel at Doppler-shifted
wavelengths of 386  and 780~nm, respectively.
\begin{figure}[tb]
    \centering
    \includegraphics[width=7.8 cm]
                                    {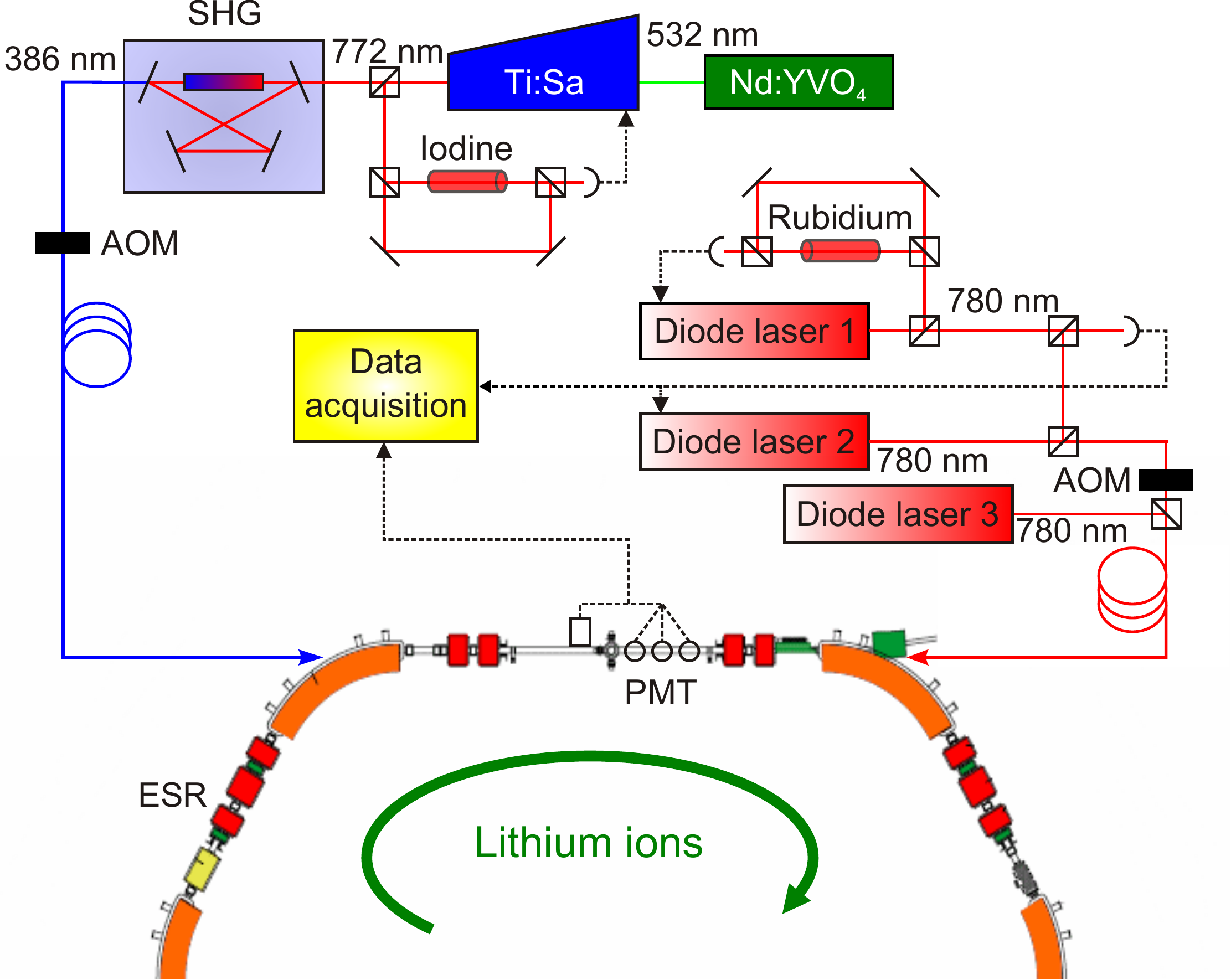}
    \caption{Left: Experimental setup. Laser system and data
             acquisition are inside a laser laboratory from which
             the light is guided to the ESR via polarization maintaining glass
             fibers.}
    \label{Bild_Aufbau}
\end{figure}
Figure~\ref{Bild_Aufbau} shows the experimental setup. The light for
parallel excitation at a fixed wavelength $\lambda_p=386$~nm
is generated by a  titanium-sapphire (Ti:Sa) laser at 772~nm, pumped
by a Nd:YVO$_4$ laser, and subsequent resonant second harmonic
generation (SHG) in a beta barium borate crystal. The Ti:Sa
laser is stabilized at 772~nm to the rovibronic P(42)1-14 transition
in molecular $^{127}$I$_2$ by frequency modulation saturation
spectroscopy~\cite{Hal81}. This transition was calibrated versus a
Rb clock using a frequency comb generator~\cite{Rei07a}. The light
for antiparallel excitation, tunable around
$\lambda_a=780$~nm, is generated by a system  of a low-power
(20~mW, diode laser 1) and a high-power (58~mW, diode laser 2)
external cavity diode laser, both in standard Littrow geometry. The
low-power laser is fixed in frequency and referenced to a transition
in $^{87}$Rb~\cite{Bar91} via frequency modulation saturation
spectroscopy. The high-power laser is stabilized to the
fixed-frequency laser via a tunable frequency-offset
lock~\cite{Sch99} allowing a tuning range of 600~MHz.
The third laser diode running at $\lambda'_a=780$~nm is
controlled by a wave meter and used to increase the signal-to-noise ratio
as discussed below.

After passing through acousto-optic frequency shifters (AOM), the
light beams at $\lambda_p$ and $\lambda_a$ are guided
together with that at $\lambda'_a$ to the ESR
via polarization maintaining fibers. The angles and
positions of the laser beams are controlled with accuracies of
$\Delta\vartheta = 20\,\mu$rad and $\Delta x = 50\,\mu$m by
motorized rotation and translation stages. Overlap within
$80\,\mu$rad between the laser beams and the ion beam is guaranteed
with vertical and horizontal scrapers at two positions along the ESR
experimental section. The fluorescence of the ions is detected by
photomultipliers (PMT)
in photon counting mode. Four types of spectra are taken applying
(i) both lasers, (ii) $\lambda_a$ only, (iii)
$\lambda_p$ only, and (iv) no laser. Using the AOMs, the
lasers are switched between these four configurations in subsequent
time windows of $100\,\mu$s duration, and the PMT signals are recorded
in four different counters gated by the AOM switching signal. This
scheme, which limits the measurement times per configuration to
$\lesssim 100$ round-trips of the ions in the ESR, allows us to
reduce background events as discussed below.

The remaining longitudinal velocity spread of the \Li ions after electron
cooling
leads to  Doppler broadening on the order of 1~GHz. To reach sub-MHz
accuracy, we single out a narrow velocity class around a central
velocity  $\beta_0$ for the Doppler shift measurements. Simultaneous
resonance of these ions with both lasers is probed to test
Eq.~(\ref{eq:Doppler1}). In our previous experiments at the
TSR~\cite{Saa03,Rei07}, we employed saturation spectroscopy on the
$\DSE (F=5/2) \,\rightarrow\, \DPZ (F=7/2)$ two-level transition and
used the Lamb dip as the observable. At the ESR the number of metastable
ions and, thus, the achievable signal-to-noise ratio turned out to be
too small to apply saturation spectroscopy. We thus employed OODR
spectroscopy on the $\Lambda$-type three-level system
within the hyperfine structure of the \Li \DSE $\! \rightarrow\!$ \DPZ line
to single out a narrow velocity class. Contrary to saturation
spectroscopy where a small Lamb dip has to be identified within a
large fluorescence background, OODR produces a positive peak on a
small background and, thus, requires less scattered photons for a
similar signal-to-noise ratio.

The $\Lambda$ system used is composed
of  the $\DSE (F = 3/2) \,\rightarrow\, \DPZ (F = 5/2)$ and $\DSE (F
= 5/2) \,\rightarrow\, \DPZ (F = 5/2)$ transitions with rest-frame
frequencies $\nu_1$ and $\nu_2$, respectively.
In the present experiment, the fixed-frequency parallel laser ($\lambda_p$)
is resonant with the $\DSE (F = 3/2) \,\rightarrow\, \DPZ (F = 5/2)$ leg of the
$\Lambda$ for a central velocity class $\beta_0$
and pumps these ions into the ground state of
the opposite leg. The second antiparallel laser ($\lambda_a$) is tuned across
this opposite
$\Lambda$ leg. All ions, except those resonant with the
fixed-frequency laser, are pumped dark after a few absorption and
emission cycles.
The $\Lambda$ resonance appears due to the continuous back-and-forth
pumping of the ions between the two ground states when the lasers
resonantly drive both legs of the $\Lambda$ for the same velocity
class $\beta_0$.

Ideally, the linewidth of the $\Lambda$ resonance should be mainly
determined by the power-broadened and time-of-flight broadened
natural linewidth. However, an ion circulating in the storage ring
experiences velocity drifts within the Doppler distribution of the
beam. Any group of ions pumped dark will, thus, get shifted back into
resonance after many round-trips in the ESR. Depending on the
velocity-drift dynamics and the frequency difference between the two
laser beams in the ions' rest frame, these ions contribute
background fluorescence and cause an additional broadening of the
line shape. To reduce background contributions due to this
drift-induced $\Lambda$ fluorescence the switching scheme mentioned
above was employed. Moreover, mainly caused by the laser-ion overlap
in the ESR-bending magnets (see Fig.~\ref{Bild_Aufbau}), the
$\Lambda$ system is not completely closed; to increase the
signal-to-noise ratio, ions in the $\DSE (F = 1/2)$ hyperfine
structure state are repumped into the $\Lambda$ system by the
antiparallel laser ($\lambda'_a$), which is tuned to the $\DSE (F =
1/2) \!\rightarrow\! \DPZ (F = 1/2)$ transition for ions at
$\beta_0$.


\begin{figure}[tb]
    \centering
        \includegraphics[width=7.8cm]
{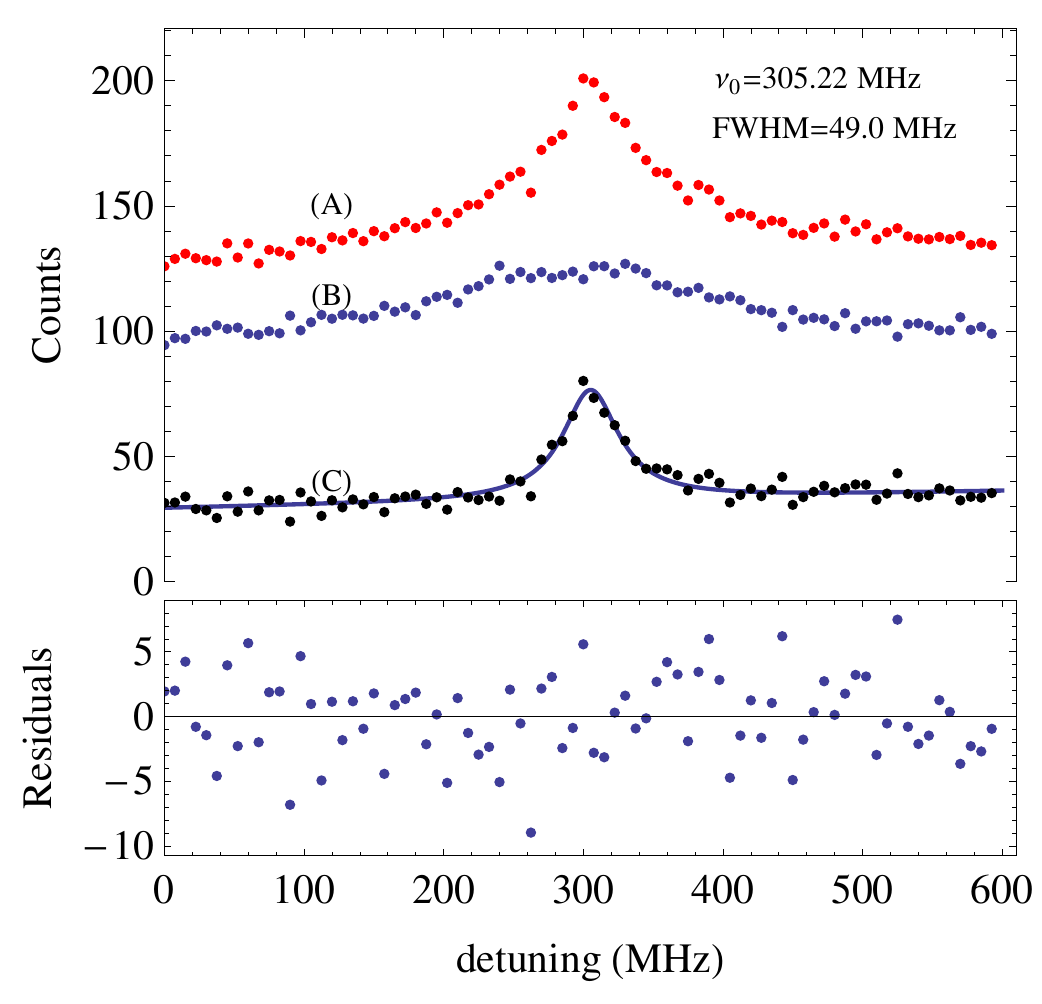}
    \caption{Typical fluorescence spectra obtained in an OODR
run containing 44
     laser scans. The scan direction is towards higher detuning. Upper graph:
($A$)
     Dark-count-rate reduced spectrum taken with all lasers on. ($B$) Spectrum
obtained by
     summing the dark-count-rate reduced spectra taken with the parallel and
antiparallel laser
     applied separately.
     Subtraction of spectrum ($B$) from ($A$) yields spectrum ($C$). The solid line
reflects the
     fit of ($C$) by a Lorentzian together with a linear background.
     Lower graph: Residuals of the fit.   }
    \label{Bild_Signal}
\end{figure}

\begin{table}[b]
\caption{Frequencies with 1$\sigma$ uncertainty budget used to determine
     $\epsilon$ by Eq.~(\ref{eq:Doppler2})  (in MHz). }
  \centering
    \begin{tabular}{l l}
      \hline\hline
   Rest frame frequencies & \hspace{0.3cm} $\Delta\nu$  \\ [0.5ex] \hline
   \hspace{0.3cm} $\nu_1=546~455~143.0 $~\cite{Rii94,Kow83}& \hspace{0.3cm} 0.43\\
   \hspace{0.3cm} $\nu_2=546~474~960.7 $~\cite{Rii94,Kow83} & \hspace{0.3cm} 0.43\\ [0.5ex]
   Doppler shifted frequencies & \hspace{0.3cm} $\Delta\nu$ \\ [0.5ex] \hline
   \multicolumn{1}{l}{Antiparallel:} \\
    \hspace{0.7cm} $\Lambda$ fit (weighted average of 40 spectra) & \hspace{0.3cm} 0.6 \\
    \hspace{0.7cm} $^{87}$Rb frequency reference & \hspace{0.3cm} 0.05 \\
    \hspace{0.7cm} Frequency stabilization diode laser& \hspace{0.3cm} 0.64 \\
    \hspace{0.7cm} Residual background linearization & \hspace{0.3cm} 0.5  \\
    \hspace{0.7cm} Gouy phase shift & \hspace{0.3cm} 1.0\\
    \hspace{0.7cm} Angular laser and ion beam alignment  & \hspace{0.3cm} 0.6\\
    \hspace{0.7cm} Ion beam divergence & \hspace{0.3cm} 0.4   \\
    \hspace{0.3cm}$\nu_a=384~225~534.98$& \hspace{0.3cm} 1.6 \\ [0.5ex]
    \multicolumn{1}{l}{Parallel:} \\
    \hspace{0.7cm} $^{127}$I$_2$ frequency reference & \hspace{0.3cm} 0.06 \\
    \hspace{0.7cm} Frequency stabilization Ti:Sa & \hspace{0.3cm} 0.12\\
    \hspace{0.7cm} Gouy phase shift &  \hspace{0.3cm} 1.0 \\
    \hspace{0.7cm} Angular laser and ion beam alignment  & \hspace{0.3cm} 0.6 \\
    \hspace{0.7cm} Ion beam divergence & \hspace{0.3cm} 0.4 \\
    \hspace{0.3cm}$\nu_p$=777~210~326.98 & \hspace{0.3cm} 1.25 \\
    \hline\hline
    \end{tabular}
 \label{Tabelle_Unsicherheiten}
\end{table}

Figure~\ref{Bild_Signal} shows a typical set of
fluorescence spectra, where the PMT dark count rates determined
in the measurement cycle (iv) have already been subtracted.
The data are averaged over 44 laser scans.
As the decay of the metastable ions causes a temporal decay of the
fluorescence signal, we decoupled the scan from the ion
injection cycle. After 69 of the 81 data points the laser scan is
paused, new ions are injected and electron-cooled, and the laser
scan is continued. Upon summing up many laser scans, the
number of metastable ions contributing to each data point is approximately
equal. Subtracting spectrum ($B$), which includes background
caused by laser stray light and contributions from the drift-induced  $\Lambda$
fluorescence, from spectrum ($A$) taken with both lasers on simultaneously,
one obtains spectrum ($C$), which displays the $\Lambda$ resonance signal
together with a smooth residual background. The fit of the residual spectrum ($C$)
by a Lorentzian together with a linear background
results in a central detuning frequency of $\nu_0 = 305.2$~MHz relative
to the iodine marker and a linewidth (FWHM) of $\delta
\nu_a = 49$~MHz.
Transformed into the \Li rest frame $\delta \nu_a$ results in a resonance width
of 35~MHz, a factor of $\sim 5$ larger than expected from the natural line widths and
the saturation as well as time-of-flight broadening.
We believe that the major part of the resonance broadening as well
as the residual background
are caused by the drift-induced $\Lambda$ fluorescence.

In total, we fitted 40 experimental $\Lambda$
spectra with Lorentzians on a linear background. The fit residuals
do not show any significant asymmetry or any other deviations from this
model and we extracted a weighted average of the frequency~$\nu_a$ of
the antiparallel laser at resonance with a statistical uncertainty
of 0.6~MHz. The slight asymmetry of the residual background
and the influence of its approximation by a straight line on the
frequency determination was simulated and is accounted for
by a systematic uncertainty of 0.5 MHz.
Table~\ref{Tabelle_Unsicherheiten} summarizes the relevant
frequencies and the uncertainty budget.
Besides the statistical error, and the uncertainties caused by the residual
background and the frequency stabilization of the laser diode,
the dominant systematic uncertainty is due to
the variation of the Gouy phases along the Gaussian laser
beams. Ions moving along the laser beam direction experience this
spatial phase variation as a frequency shift~\cite{Rei07}. From an
experimentally validated numerical simulation of this effect we
infer a correction of $\nu_\text{Ph} =0.44\pm 1.0$~MHz for each
laser, using measured laser beam parameters. Further systematic
uncertainties arise from the residual divergence of the
electron-cooled ion beam 
as well as from a possible angular misalignment
between laser and ion
beams.

Inserting the rest-frame frequencies and the Doppler shifted frequencies into Eq.~(\ref{eq:Doppler2})
we find
\begin{equation}
    \epsilon(\beta)=\sqrt{\frac{\nu_a \nu_p}{\nu_1\nu_2}} - 1 =
    (1.5 \pm 2.3)  \times 10^{-9},
    \label{Gleichung_AlphaAusFrequenzen}
\end{equation}
at $\beta=0.338$. {From the experimental uncertainty, we find} an
upper limit for the reduced IS-observable of $|\alpha|
=|\epsilon(\beta)/\beta^2| \leq 2.0 \times 10^{-8}$. This is not
only a four-fold improvement over previous IS experiments~\cite{Rei07},
but in contrast to Ref.~\cite{Rei07}, where the measurement of the rest
frequency was replaced by a measurement at low ion velocity assuming
a polynomial dependence up to second order for
$\epsilon(\beta)$, the present result is also independent of any
assumption about the velocity dependence of $\epsilon(\beta)$.

When interpreted within the kinematic Robertson-Mansouri-Sexl (RMS)
test theory~\cite{Rob49,Man77,Kre92,Wil92}, the IS experiment,
together with the Michelson-Morley~\cite{Mic87} and the
Kennedy-Thorndike~\cite{Ken32} experiments, belongs to the three
classic tests of SR and the reduced observable $\alpha$ coincides with the RMS parameter
$\hat\alpha^{\rm{(RMS)}}$~\cite{Nov09}. In this model,
$\hat\alpha^{\rm{(RMS)}}$ is also accessible by experiments testing the
anisotropy of the speed of light. The most accurate test of this
kind used macroscopic clocks of the global positioning system and
found a limit $|\hat\alpha^{\rm{(RMS)}}|<1\times 10^{-6}$~\cite{Wolf97},
50 times less sensitive than our experiment.

A dynamical test framework allowing for particle LI violations
is provided by the Standard Model
Extension (SME)~\cite{Kos89,Kos02}. It is an effective field theory
that extends the Standard Model Lagrangian density by Lorentz
violating terms containing operators of arbitrary mass dimension.
Within the hitherto worked out minimal SME, which is restricted to
mass dimension $\leq 4$, the IS experiment provides absolute bounds
on proton and electron
parameters~\cite{Lan05,Kos11} as well as on the isotropy parameter
in the photon sector~\cite{Tob05} at a boost velocity
of $0.34$~c. We find $|\tilde c_Q^{p}|<2\times
10^{-11}$ for the proton and $|\tilde c_Q^{e}|<2\times 10^{-8}$ for
the electron~\cite{footnote}, both by many orders of magnitude less
sensitive than $Cs$ fountain clock \cite{Wolf06} and microwave
resonator \cite{Mueller07} experiments, respectively, which, however,
measure at velocities connected with
sidereal variations. In the photon sector of the minimal SME we find
a constraint for the isotropy parameter of $|\tilde\kappa_{tr}|\leq
2\times 10^{-8}$. This is a factor of 20 short of the best direct
test~\cite{Bay12}. Other model-dependent constraints of these model
parameters come from
various experiments on high-energy cosmic rays~\cite{Kli08} ($10^{-20}$ level),
relativistic electrons~\cite{Altschul09,Hohensee09}, and
contributions to the anomalous magnetic moment of the
electron~\cite{Car06}.

As several systematic uncertainties are in the same order of
magnitude and each of them difficult to improve, the constraint of
Eq.~(\ref{Gleichung_AlphaAusFrequenzen}) constitutes the concluding
result of a generation of storage-ring based Ives-Stilwell
experiments.  While its sensitivity to the minimal SME parameters is
lower than those of astrophysical observations and interferometer
experiments, both searching for sidereal variations, the IS
experiment stands out as one of the few absolute measurements being
independent from sidereal variations by using an experimentally
prepared large Lorentz boost. {For a putative effect that scales
with $\beta^4$, our result sets a 100-fold stronger limit than
previous measurements~\cite{Nov09,Rei07,MacArthur}}. The full
benefit of this boost may only become visible when analyzed in the
full SME including operators of higher mass dimension. This
challenging theoretical work has only begun~\cite{Kos13}.


This work was supported by the German Federal Ministry of Education and
Research (BMBF, Contract No. 06MZ9179I),
the Helmholtz Association (Contract No. VH-NG-148), and
the Deutsche Forschungsgemeinschaft (DFG, Contract No. NO789/1-1).
G.G. acknowledges support by NSERC (Canada). A. K.
acknowledges support from the Carl-Zeiss-Stiftung
(Grant No. AZ:21-0563-2.8/197/1). D.S. acknowledges support by the
Weizmann Institute of Science through the Joseph Meyerhoff program.


\begin{thebibliography}{99}
\bibitem{footnote}
            {Values quoted in \cite{Lan05} for the
            TSR experiments are by a factor of 10 too low.}
\end{thebibliography}
\end{document}